\documentclass{article}

\usepackage{PRIMEarxiv}

\usepackage[utf8]{inputenc} 
\usepackage[T1]{fontenc}    
\usepackage{hyperref}       
\usepackage{url}            
\usepackage{booktabs}       
\usepackage{amsfonts}       
\usepackage{amsmath,amssymb}
\usepackage{mathtools}
\usepackage{nicefrac}       
\usepackage{microtype}      
\usepackage{lipsum}
\usepackage{fancyhdr}       
\usepackage{graphicx}       
\usepackage{amsthm}
\usepackage{times}
\usepackage{subfig}
\usepackage{xcolor}
\usepackage{array}
\usepackage{cite}
\usepackage{multirow}
\usepackage{adjustbox}
\usepackage[scr=dutchcal]{mathalfa}
\usepackage[font=small,labelfont=bf]{caption}
\usepackage{enumitem}
\usepackage{tikz}
\usetikzlibrary{arrows,matrix} 

\newtheorem{theorem}{Theorem}

\newtheorem{proposition}{Proposition}

\newtheorem{assumption}{Assumption}

\usepackage{tikz-network}
\tikzset{
    block/.style = {draw, fill=white, rectangle, minimum height=.75cm, minimum width=1.25cm},
    Circle/.style = {draw, thick, fill=white, circle, minimum height=.75cm, minimum width=1.25cm},
    tmp/.style  = {coordinate}, 
    sum/.style= {draw=white, fill=white, circle, minimum height=.75cm, minimum width=1.25cm},
    input/.style = {coordinate},
    output/.style= {coordinate},
    pinstyle/.style = {pin edge={to-,thin,black}
    }
}

\title{Barrier-Based Safety Filtering for Quadrotor Flight in Cluttered Environments Under Acceleration Limits}

\author{
  Andi Rahmadiansah \\
  Department of Engineering Physics \\
  Institut Teknologi Sepuluh Nopember \\
  Surabaya 60111, Indonesia \\
  \texttt{andi.rahmadiansah@its.ac.id}
}

\begin{document}
\maketitle

\begin{abstract}
Accurate trajectory tracking alone does not guarantee safe quadrotor flight
when the reference passes near obstacles or the available control authority is
limited. This paper presents a safety-filtered trajectory-tracking framework
for a quadrotor operating at a fixed altitude in obstacle-cluttered
environments. A nominal tracking controller is combined with high-order
control barrier constraints and an acceleration-constrained quadratic
program. The safety filter preserves the nominal acceleration whenever it is
admissible and modifies it only when the prescribed obstacle clearance is
threatened. The analysis establishes nominal exponential tracking, forward
invariance of the safe region under compatible initial conditions and
continued quadratic-program feasibility, and bounded tracking error during
safety intervention. Numerical experiments using an idealized planar translational model
with single and multiple stationary circular obstacles show minimum
clearances of $0.000$--$0.001$~m under the safety filter, compared with
$-1.594/-1.421$~m under nominal control. Tracking RMSE increases from
$0.618$ to $0.950$~m across reference-speed scales $0.6$--$1.6$,
illustrating the effect of reference speed on tracking accuracy
under safety constraints.
Additional tests examine how acceleration limits affect control effort
and feasibility. These findings inform safety-filter design for UAV
trajectory tracking under limited control authority.
\end{abstract}
\allowdisplaybreaks


\section{Introduction}
\label{sec:introduction}

Quadrotor unmanned aerial vehicles (UAVs) are increasingly used for
inspection, surveillance, mapping, infrastructure monitoring, and navigation
in confined environments
\cite{ref-1,ref-2,ref-3,ref-4}. Their maneuverability and
ability to hover make them attractive for operation near buildings and other
structures, but these applications require the vehicle to follow a prescribed
trajectory while maintaining sufficient separation from surrounding
obstacles \cite{ref-5,ref-6}. Although conventional
trajectory-tracking controllers can provide accurate motion in an
obstacle-free environment \cite{ref-7,ref-8,ref-9}, they do not explicitly
enforce collision-avoidance requirements. A reference trajectory that
approaches an obstacle, an initial tracking error, or limited control
authority may therefore lead to a collision even when the nominal controller
performs as intended.

Obstacle avoidance can be addressed at either the planning or control level.
Sampling-based, graph-search, and optimization-based planners can generate
collision-free geometric routes before flight
\cite{ref-10,ref-11}. However, the resulting route may become
unsafe during execution because of tracking errors, disturbances, inaccurate
obstacle information, and modeling uncertainty \cite{ref-12,ref-13,ref-14}.
Reactive methods, including artificial potential fields, modify the vehicle
motion using the current obstacle configuration
\cite{ref-15,ref-16}. Although computationally attractive, their behavior
can depend strongly on parameter selection and may exhibit local minima,
oscillatory motion, or conflicting attractive and repulsive actions
\cite{ref-17}. These observations motivate a safety layer that operates
directly on the control command while preserving the original tracking
controller whenever possible.

Control barrier functions provide a systematic method for enforcing
state-dependent safety constraints
\cite{ref-18,ref-19}. A nominal control command can be passed through
a quadratic-programming safety filter that computes the closest command
satisfying the barrier inequalities \cite{ref-20,ref-21}. Consequently,
tracking performance and safety can be designed separately: the nominal
controller determines the desired flight behavior, whereas the barrier
constraints prevent the system from entering an unsafe region. Robust,
adaptive, stochastic, and learning-based extensions have also been developed
to account for uncertain dynamics, noisy measurements, and incomplete
environmental information \cite{ref-22,ref-23}. This modular
structure is particularly attractive for quadrotors because the safety filter
can be added to an existing outer-loop tracking controller without requiring
a complete redesign of the inner-loop attitude dynamics
\cite{ref-24,ref-25}.

A practical difficulty arises because a quadrotor cannot generate arbitrary
acceleration. Barrier constraints that are feasible under unlimited control
authority may become infeasible after acceleration or thrust limits are
imposed. This issue is especially important when the vehicle moves rapidly
toward an obstacle, begins near the boundary of the safe region, or encounters
several obstacles simultaneously. Recent studies have therefore investigated
barrier-function compatibility, compositions of multiple barrier
constraints, constrained-input feasibility, and recursively feasible safety
filters \cite{ref-26,ref-27}. When the safety function depends only on
position and the control input is acceleration, the constraint has relative
degree two. High-order or exponential control barrier functions are then
required to express the safety condition as an inequality involving the
control input \cite{ref-28}. A useful evaluation of safety-filtered tracking
should consequently consider not only obstacle clearance, but also tracking
error, control effort, reference speed, acceleration saturation, and
continued feasibility of the constrained optimization problem.

This paper studies safety-filtered trajectory tracking for a quadrotor flying
at a fixed altitude in a planar obstacle environment. The outer-loop
translational dynamics are represented by a double-integrator model, while a
proportional-derivative controller provides the nominal acceleration command.
Circular obstacles are incorporated through high-order control barrier
constraints because the acceleration input appears only after differentiating
the position-based safety function twice. The applied acceleration is then
obtained from a quadratic program that enforces both obstacle-separation
requirements and componentwise acceleration limits.

The main contributions are summarized as follows:
\begin{enumerate}
    \item A modular safety-filtered trajectory-tracking framework is formulated
    for a planar quadrotor under explicit acceleration constraints. The
    framework preserves the nominal controller whenever its command satisfies
    the safety and input constraints.

    \item The safety and tracking properties of the closed-loop system are
    characterized. Forward invariance of the safe region is established under
    compatible initial barrier conditions and continued quadratic-program
    feasibility.

    \item The tracking-error dynamics are expressed as a stable nominal system
    driven by the modification introduced by the safety filter. This
    representation clarifies the tradeoff between accurate reference tracking
    and obstacle avoidance.

    \item A systematic numerical evaluation examines single and multiple
    obstacles and quantifies the effects of acceleration limits and reference
    speed on tracking error, obstacle clearance, control effort, safety-filter
    activity, and optimization feasibility.
\end{enumerate}

The results show that small nominal tracking error does not necessarily imply
safe flight. They also demonstrate that insufficient acceleration can render
the barrier constraints infeasible, whereas increasing reference speed
requires more frequent and more energetic safety interventions.

The remainder of this paper is organized as follows.
Section~\ref{sec:problem} introduces the planar quadrotor model, obstacle
representation, and tracking objective. Section~\ref{sec:safety_filter}
develops the high-order barrier constraints and acceleration-constrained
safety filter. Section~\ref{sec:analysis} presents the safety, tracking, and
feasibility analysis. Section~\ref{sec:numerical} reports the numerical
experiments and discusses the observed performance tradeoffs. Finally,
Section~\ref{sec:conclusion} concludes the paper.

\section{Quadrotor Model and Problem Formulation}
\label{sec:problem}

\subsection{Planar Quadrotor Motion}

Consider a quadrotor flying at a fixed altitude in a planar environment. Its
horizontal position and velocity are denoted by
\begin{equation}
    p=
    \begin{bmatrix}
        x & y
    \end{bmatrix}^{\top}
    \in\mathbb{R}^{2},
    \qquad
    v=
    \begin{bmatrix}
        v_x & v_y
    \end{bmatrix}^{\top}
    \in\mathbb{R}^{2}.
\end{equation}
The outer-loop translational dynamics are represented by
\begin{equation}
    \dot p=v,
    \qquad
    \dot v=u,
    \label{eq:outer_loop_model}
\end{equation}
where
\begin{equation}
    u=
    \begin{bmatrix}
        u_x & u_y
    \end{bmatrix}^{\top}
    \in\mathbb{R}^{2}
\end{equation}
is the commanded horizontal acceleration.

The altitude is maintained at a prescribed value,
\begin{equation}
    z(t)=z_0,
\end{equation}
by a sufficiently fast vertical and attitude controller. The horizontal
acceleration command $u$ is converted into the corresponding roll, pitch, and
thrust commands by the inner-loop controller. The model
\eqref{eq:outer_loop_model} is therefore used for the trajectory-tracking and
safety-filter design.

The available horizontal acceleration is limited. Define the admissible input
set
\begin{equation}
    \mathcal{U}
    :=
    \left\{
        u\in\mathbb{R}^{2}:
        u_{\min}\leq u\leq u_{\max}
    \right\},
    \label{eq:admissible_acceleration}
\end{equation}
where the inequalities are interpreted componentwise. For symmetric
acceleration limits,
\begin{equation}
    u_{\min}=-u_{\max}.
\end{equation}

\subsection{Nominal Trajectory-Tracking Controller}

Let $p_d:[0,\infty)\rightarrow\mathbb{R}^{2}$ be a twice continuously
differentiable reference trajectory. Define the position and velocity
tracking errors as
\begin{equation}
    e_p=p-p_d,
    \qquad
    e_v=v-\dot p_d.
    \label{eq:tracking_errors}
\end{equation}
In the absence of obstacles, consider the nominal acceleration command
\begin{equation}
    u_{\mathrm{nom}}
    =
    \ddot p_d
    -
    K_p e_p
    -
    K_v e_v,
    \label{eq:nominal_controller}
\end{equation}
where $K_p,K_v\in\mathbb{R}^{2\times2}$ are positive-definite gain matrices.
For diagonal choices,
\begin{equation}
    K_p
    =
    \operatorname{diag}
    \{k_{p,x},k_{p,y}\},
    \qquad
    K_v
    =
    \operatorname{diag}
    \{k_{v,x},k_{v,y}\},
\end{equation}
the horizontal directions can be tuned independently.

Substituting \eqref{eq:nominal_controller} into
\eqref{eq:outer_loop_model} gives
\begin{equation}
    \dot e_p=e_v,
    \qquad
    \dot e_v
    =
    -K_p e_p-K_v e_v.
    \label{eq:nominal_error_dynamics}
\end{equation}
Thus, the nominal controller provides exponential trajectory tracking when
its command can be applied directly. However, it does not account for
obstacles and may drive the quadrotor through an unsafe region.

\subsection{Obstacle and Safety Model}

Consider $M$ stationary circular obstacles. Obstacle $j$ is characterized by
its center $p_{o,j}\in\mathbb{R}^{2}$ and physical radius $r_{o,j}>0$. The
horizontal footprint of the quadrotor is conservatively enclosed by a circle
of radius $r_{\mathrm{UAV}}>0$.

To account for the obstacle size, vehicle size, and an additional safety
margin $d_{\mathrm{safe}}>0$, define
\begin{equation}
    d_j
    =
    r_{o,j}
    +
    r_{\mathrm{UAV}}
    +
    d_{\mathrm{safe}}.
    \label{eq:required_clearance}
\end{equation}
The safety function associated with obstacle $j$ is
\begin{equation}
    h_j(p)
    =
    \|p-p_{o,j}\|_2^2-d_j^2.
    \label{eq:safety_function}
\end{equation}
The corresponding safe set is
\begin{equation}
    \mathcal{C}_j
    :=
    \left\{
        (p,v)\in\mathbb{R}^{4}:
        h_j(p)\geq0
    \right\}.
    \label{eq:individual_safe_set}
\end{equation}
Hence, $h_j(p)=0$ represents the effective obstacle boundary,
$h_j(p)>0$ represents a safe position, and $h_j(p)<0$ indicates a violation
of the required clearance.

The overall safe region is
\begin{equation}
    \mathcal{C}
    :=
    \bigcap_{j=1}^{M}\mathcal{C}_j.
    \label{eq:overall_safe_set}
\end{equation}
The required collision-avoidance condition is
\begin{equation}
    \|p(t)-p_{o,j}\|_2
    \geq
    d_j,
    \qquad
    j=1,\ldots,M,
    \quad
    t\geq0.
    \label{eq:safety_requirement}
\end{equation}

The following assumptions are adopted.

\begin{assumption}
\label{ass:reference}
The reference trajectory $p_d(t)$ is twice continuously differentiable, and
$p_d(t)$, $\dot p_d(t)$, and $\ddot p_d(t)$ are bounded.
\end{assumption}

\begin{assumption}
\label{ass:obstacles}
The obstacle centers and radii are known and remain constant during the
flight. The horizontal position and velocity of the quadrotor are available
to the safety filter.
\end{assumption}

\begin{assumption}
\label{ass:inner_loop}
The altitude and attitude controllers are sufficiently fast that the
horizontal dynamics can be represented by \eqref{eq:outer_loop_model}.
\end{assumption}

\subsection{Control Objective}

The nominal command \eqref{eq:nominal_controller} prioritizes trajectory
tracking but does not guarantee \eqref{eq:safety_requirement}. The objective
is to construct an online safety filter that generates an applied acceleration
$u^\star$ satisfying the following requirements:
\begin{enumerate}
    \item $u^\star(t)\in\mathcal{U}$ for all $t\geq0$;
    \item the safe set $\mathcal{C}$ remains forward invariant under
    appropriate initial safety conditions;
    \item $u^\star$ remains as close as possible to $u_{\mathrm{nom}}$;
    \item whenever the nominal command is safe and admissible,
    $u^\star=u_{\mathrm{nom}}$.
\end{enumerate}

The safety filter should therefore preserve the nominal tracking behavior
when no obstacle presents an immediate risk and modify the horizontal
acceleration only when required to maintain safety. The next section
constructs this filter using high-order control barrier conditions and a
constrained quadratic program.

\section{Safety-Filtered Trajectory-Tracking Control}
\label{sec:safety_filter}

\subsection{High-Order Control Barrier Condition}

The safety function $h_j(p)$ in \eqref{eq:safety_function} depends only on
position, whereas the control input $u$ appears in the acceleration dynamics.
Consequently, the input does not appear in the first derivative of $h_j$.
Differentiating along \eqref{eq:outer_loop_model} gives
\begin{equation}
    \dot h_j(p,v)
    =
    2(p-p_{o,j})^\top v.
    \label{eq:safety_function_first_derivative}
\end{equation}
Differentiating once more yields
\begin{equation}
    \ddot h_j(p,v,u)
    =
    2\|v\|_2^2
    +
    2(p-p_{o,j})^\top u.
    \label{eq:safety_function_second_derivative}
\end{equation}
Thus, $h_j$ has relative degree two with respect to the acceleration input.

To construct a high-order control barrier condition, define
\begin{equation}
    \psi_{j,0}
    :=
    h_j
\end{equation}
and
\begin{equation}
    \psi_{j,1}
    :=
    \dot h_j+\alpha_1 h_j,
    \label{eq:first_barrier_function}
\end{equation}
where $\alpha_1>0$. The second barrier condition is imposed as
\begin{equation}
    \dot\psi_{j,1}
    +
    \alpha_2\psi_{j,1}
    \geq0,
    \label{eq:second_barrier_condition}
\end{equation}
where $\alpha_2>0$. Substituting
\eqref{eq:first_barrier_function} into
\eqref{eq:second_barrier_condition} gives
\begin{equation}
    \ddot h_j
    +
    (\alpha_1+\alpha_2)\dot h_j
    +
    \alpha_1\alpha_2h_j
    \geq0.
    \label{eq:high_order_barrier_condition}
\end{equation}

Using \eqref{eq:safety_function_first_derivative} and
\eqref{eq:safety_function_second_derivative}, condition
\eqref{eq:high_order_barrier_condition} becomes
\begin{equation}
\begin{split}
    2(p-p_{o,j})^\top u
    \geq{}&
    -2\|v\|_2^2\\
    &-2(\alpha_1+\alpha_2)
    (p-p_{o,j})^\top v\\
    &-\alpha_1\alpha_2h_j(p).
\end{split}
\label{eq:affine_barrier_constraint}
\end{equation}
The barrier condition is therefore affine in the decision variable $u$.

For compactness, define
\begin{equation}
    a_j(p)
    :=
    2(p-p_{o,j})
    \in\mathbb{R}^{2}
\end{equation}
and
\begin{equation}
\begin{split}
    b_j(p,v)
    :={}&
    -2\|v\|_2^2\\
    &-2(\alpha_1+\alpha_2)
    (p-p_{o,j})^\top v\\
    &-\alpha_1\alpha_2h_j(p).
\end{split}
\end{equation}
Then \eqref{eq:affine_barrier_constraint} can be written as
\begin{equation}
    a_j(p)^\top u
    \geq
    b_j(p,v).
    \label{eq:compact_barrier_constraint}
\end{equation}

The high-order safe set associated with obstacle $j$ is
\begin{equation}
\begin{split}
    \mathcal{C}_{j}^{\mathrm{H}}
    :=
    \bigl\{
        (p,v)\in\mathbb{R}^{4}:
        \;&h_j(p)\geq0,\\
        &\psi_{j,1}(p,v)\geq0
    \bigr\}.
\end{split}
\label{eq:high_order_safe_set}
\end{equation}
The additional condition $\psi_{j,1}\geq0$ restricts how rapidly the
quadrotor may approach an obstacle. In particular, a position may be outside
the obstacle while the inward velocity is too large to satisfy the required
high-order safety condition.

\subsection{Acceleration-Constrained Safety Filter}

The applied acceleration is selected to remain as close as possible to the
nominal tracking command while satisfying the barrier and actuator
constraints. At each time instant, consider the quadratic program
\begin{equation}
\begin{aligned}
    u^\star
    =
    \arg\min_{u\in\mathbb{R}^{2}}
    \quad&
    \frac{1}{2}
    (u-u_{\mathrm{nom}})^\top
    W
    (u-u_{\mathrm{nom}})
    \\
    \text{subject to}\quad&
    a_j(p)^\top u
    \geq
    b_j(p,v),
    \quad
    j\in\mathcal{J}_a,
    \\
    &
    u_{\min}\leq u\leq u_{\max},
\end{aligned}
\label{eq:safety_filter_qp}
\end{equation}
where $W\in\mathbb{R}^{2\times2}$ is positive definite and
$\mathcal{J}_a$ denotes the set of active obstacles. The identity choice
$W=I_2$ penalizes acceleration modifications equally in both horizontal
directions.

The optimization problem \eqref{eq:safety_filter_qp} is convex because its
objective is quadratic and its constraints are affine. Hence, whenever its
feasible set is nonempty, it admits a unique minimizer. If
$u_{\mathrm{nom}}$ satisfies all barrier and acceleration constraints, then
\begin{equation}
    u^\star=u_{\mathrm{nom}}.
\end{equation}
The filter therefore leaves the nominal tracking controller unchanged when
its command is safe and admissible.

Define the safety-induced acceleration modification as
\begin{equation}
    \Delta u
    :=
    u^\star-u_{\mathrm{nom}}.
    \label{eq:safety_modification}
\end{equation}
Substitution into the tracking-error dynamics gives
\begin{equation}
    \dot e_p=e_v,
    \qquad
    \dot e_v
    =
    -K_p e_p-K_v e_v+\Delta u.
    \label{eq:filtered_error_dynamics}
\end{equation}
Thus, the safety filter can be interpreted as an additional input acting on
the nominal exponentially stable tracking dynamics.

\subsection{Single-Obstacle Interpretation}

The action of the safety filter can be seen directly when only one obstacle
constraint is active and the acceleration bounds are temporarily omitted.
Let
\begin{equation}
    a=a_j(p),
    \qquad
    b=b_j(p,v).
\end{equation}
For $W=I_2$, the safety filter becomes
\begin{equation}
\begin{aligned}
    \min_{u\in\mathbb{R}^{2}}
    \quad&
    \frac{1}{2}\|u-u_{\mathrm{nom}}\|_2^2
    \\
    \text{subject to}\quad&
    a^\top u\geq b.
\end{aligned}
\end{equation}
Its solution is
\begin{equation}
    u^\star
    =
    \begin{cases}
        u_{\mathrm{nom}},
        &
        a^\top u_{\mathrm{nom}}\geq b,
        \\[2mm]
        \displaystyle
        u_{\mathrm{nom}}
        +
        \frac{
            b-a^\top u_{\mathrm{nom}}
        }{
            \|a\|_2^2
        }a,
        &
        a^\top u_{\mathrm{nom}}<b.
    \end{cases}
    \label{eq:single_obstacle_solution}
\end{equation}

Since $a=2(p-p_{o,j})$ points from the obstacle center toward the quadrotor,
the correction in \eqref{eq:single_obstacle_solution} acts in the outward
direction. The nominal controller is modified only when it violates the
barrier condition, and the modification is the smallest Euclidean correction
that restores the constraint.

\subsection{Active Obstacle Selection}

Including every obstacle in \eqref{eq:safety_filter_qp} is unnecessary when
many obstacles are far from the quadrotor. Define the distance from the
quadrotor center to the effective boundary of obstacle $j$ as
\begin{equation}
    \delta_j(p)
    :=
    \|p-p_{o,j}\|_2-d_j.
    \label{eq:obstacle_boundary_distance}
\end{equation}
Given an activation distance $d_{\mathrm{act}}>0$, the active obstacle set is
selected as
\begin{equation}
    \mathcal{J}_a(p)
    :=
    \left\{
        j\in\{1,\ldots,M\}:
        \delta_j(p)\leq d_{\mathrm{act}}
    \right\}.
    \label{eq:active_obstacle_set}
\end{equation}
Only the obstacles in $\mathcal{J}_a(p)$ contribute constraints to the
quadratic program.

The activation distance should be selected sufficiently large that an
obstacle becomes active before the high-order compatibility condition
$\psi_{j,1}\geq0$ can be violated. In the theoretical analysis, it is assumed
that all obstacles that may affect safety are included in
$\mathcal{J}_a(p)$. Setting
\begin{equation}
    \mathcal{J}_a
    =
    \{1,\ldots,M\}
\end{equation}
recovers the full safety filter without obstacle selection.

\subsection{Online Implementation}

At each sampling instant, the safety-filtered controller performs the
following steps:
\begin{enumerate}
    \item Measure or estimate the current position $p$ and velocity $v$.
    \item Evaluate the reference trajectory and compute
    $u_{\mathrm{nom}}$ from \eqref{eq:nominal_controller}.
    \item Determine the active obstacle set using
    \eqref{eq:active_obstacle_set}.
    \item Construct the affine barrier constraints
    \eqref{eq:compact_barrier_constraint}.
    \item Solve the quadratic program \eqref{eq:safety_filter_qp}.
    \item Apply $u^\star$ through the quadrotor inner-loop controller.
\end{enumerate}

Unlike the A* planner, the proposed safety filter operates online. It does not
first construct a complete obstacle-free path. Instead, it repeatedly checks
whether the nominal tracking command is safe and modifies the current
acceleration only when necessary. Safety and tracking properties of the
closed-loop system are established in the next section.

\section{Safety and Tracking Analysis}
\label{sec:analysis}

This section establishes the nominal tracking property, forward invariance of
the obstacle-free region, and boundedness of the tracking error during safety
intervention. The effect of acceleration limits on feasibility is also
discussed.

\subsection{Nominal Tracking Performance}

We first recall the behavior of the tracking controller when the safety filter
is inactive.

\begin{proposition}
\label{prop:nominal_tracking}
Suppose that $K_p$ and $K_v$ are diagonal matrices with strictly positive
diagonal entries. Under the nominal controller
\eqref{eq:nominal_controller}, the equilibrium
\begin{equation}
    e_p=0,
    \qquad
    e_v=0
\end{equation}
of \eqref{eq:nominal_error_dynamics} is globally exponentially stable.
\end{proposition}

\begin{proof}
For each horizontal direction $i\in\{x,y\}$, the tracking error satisfies
\begin{equation}
    \ddot e_{p,i}
    +
    k_{v,i}\dot e_{p,i}
    +
    k_{p,i}e_{p,i}
    =
    0.
\end{equation}
The corresponding characteristic polynomial is
\begin{equation}
    s^2+k_{v,i}s+k_{p,i}.
\end{equation}
Since $k_{p,i}>0$ and $k_{v,i}>0$, both roots have negative real parts.
Therefore, the tracking errors in both directions converge exponentially to
zero.
\end{proof}

Proposition~\ref{prop:nominal_tracking} describes the performance objective
that the safety filter should preserve whenever the nominal command is safe.

\subsection{Forward Invariance of the Safe Set}

The next result establishes collision avoidance under the high-order barrier
conditions.

\begin{theorem}
\label{thm:forward_invariance}
Suppose that the quadratic program \eqref{eq:safety_filter_qp} remains
feasible for all $t\geq0$ and that every obstacle relevant to safety is
included in $\mathcal{J}_a(p)$. If, for every obstacle $j$,
\begin{equation}
    h_j(p(0))\geq0,
    \qquad
    \psi_{j,1}(p(0),v(0))\geq0,
    \label{eq:compatible_initial_conditions}
\end{equation}
then the control input $u^\star$ generated by
\eqref{eq:safety_filter_qp} guarantees
\begin{equation}
    h_j(p(t))\geq0,
    \qquad
    t\geq0.
\end{equation}
Consequently,
\begin{equation}
    \|p(t)-p_{o,j}\|_2\geq d_j
\end{equation}
for all $j=1,\ldots,M$ and $t\geq0$.
\end{theorem}

\begin{proof}
By construction, the solution of \eqref{eq:safety_filter_qp} satisfies
\begin{equation}
    \dot\psi_{j,1}+\alpha_2\psi_{j,1}\geq0.
\end{equation}
Equivalently,
\begin{equation}
    \dot\psi_{j,1}\geq-\alpha_2\psi_{j,1}.
\end{equation}
The comparison principle gives
\begin{equation}
    \psi_{j,1}(t)
    \geq
    e^{-\alpha_2t}\psi_{j,1}(0).
\end{equation}
Since $\psi_{j,1}(0)\geq0$, it follows that
$\psi_{j,1}(t)\geq0$ for all $t\geq0$.

Using the definition
\begin{equation}
    \psi_{j,1}
    =
    \dot h_j+\alpha_1h_j,
\end{equation}
we obtain
\begin{equation}
    \dot h_j\geq-\alpha_1h_j.
\end{equation}
A second application of the comparison principle yields
\begin{equation}
    h_j(t)
    \geq
    e^{-\alpha_1t}h_j(0).
\end{equation}
Because $h_j(0)\geq0$, the inequality $h_j(t)\geq0$ holds for all
$t\geq0$. Therefore, the required obstacle clearance is preserved.
\end{proof}

Theorem~\ref{thm:forward_invariance} requires both conditions in
\eqref{eq:compatible_initial_conditions}. The position condition
$h_j(p(0))\geq0$ ensures that the quadrotor initially lies outside the
effective obstacle. The condition $\psi_{j,1}(p(0),v(0))\geq0$ additionally
ensures that its inward velocity is compatible with the selected barrier
parameters.

\subsection{Tracking Error under Safety Intervention}

Define the combined tracking-error state
\begin{equation}
    e
    :=
    \begin{bmatrix}
        e_p\\
        e_v
    \end{bmatrix}
    \in\mathbb{R}^{4}.
\end{equation}
The filtered error dynamics \eqref{eq:filtered_error_dynamics} can be written
as
\begin{equation}
    \dot e
    =
    A_e e+B_e\Delta u,
    \label{eq:compact_filtered_error}
\end{equation}
where
\begin{equation}
    A_e
    =
    \begin{bmatrix}
        0&I_2\\
        -K_p&-K_v
    \end{bmatrix},
    \qquad
    B_e
    =
    \begin{bmatrix}
        0\\
        I_2
    \end{bmatrix}.
\end{equation}
Under the conditions of Proposition~\ref{prop:nominal_tracking}, $A_e$ is
Hurwitz.

\begin{proposition}
\label{prop:tracking_bound}
Suppose that the safety-induced modification is bounded as
\begin{equation}
    \|\Delta u(t)\|_2\leq\overline{\Delta}_u,
    \qquad
    t\geq0.
    \label{eq:bounded_safety_modification}
\end{equation}
Then the tracking-error dynamics \eqref{eq:compact_filtered_error} are
input-to-state stable with respect to $\Delta u$. In particular, the tracking
error satisfies
\begin{equation}
    \limsup_{t\rightarrow\infty}\|e(t)\|_2
    \leq
    \gamma_e\overline{\Delta}_u
    \label{eq:ultimate_tracking_bound}
\end{equation}
for some constant $\gamma_e>0$.
\end{proposition}

\begin{proof}
Let $Q=Q^\top>0$ be arbitrary, and let $P=P^\top>0$ be the unique solution of
\begin{equation}
    A_e^\top P+PA_e=-Q.
    \label{eq:lyapunov_equation}
\end{equation}
Consider
\begin{equation}
    V(e)=e^\top Pe.
\end{equation}
Along \eqref{eq:compact_filtered_error},
\begin{equation}
\begin{split}
    \dot V
    &=
    -e^\top Qe
    +
    2e^\top PB_e\Delta u\\
    &\leq
    -\lambda_{\min}(Q)\|e\|_2^2
    +
    2\|PB_e\|_2\|e\|_2\|\Delta u\|_2.
\end{split}
\end{equation}
Using Young's inequality gives
\begin{equation}
\begin{split}
    \dot V
    \leq{}&
    -\frac{\lambda_{\min}(Q)}{2}\|e\|_2^2
    +
    \frac{2\|PB_e\|_2^2}
    {\lambda_{\min}(Q)}
    \|\Delta u\|_2^2.
\end{split}
\label{eq:lyapunov_iss_bound}
\end{equation}
Since
\begin{equation}
    \lambda_{\min}(P)\|e\|_2^2
    \leq
    V(e)
    \leq
    \lambda_{\max}(P)\|e\|_2^2,
\end{equation}
standard comparison arguments imply input-to-state stability. One admissible
value of the gain in \eqref{eq:ultimate_tracking_bound} is
\begin{equation}
    \gamma_e
    =
    \frac{2\|PB_e\|_2}
    {\lambda_{\min}(Q)}
    \sqrt{
        \frac{\lambda_{\max}(P)}
        {\lambda_{\min}(P)}
    }.
    \label{eq:tracking_gain}
\end{equation}
\end{proof}

Proposition~\ref{prop:tracking_bound} formalizes the safety--tracking tradeoff.
A larger modification of the nominal acceleration may be required when the
reference trajectory passes close to an obstacle. This intervention increases
the tracking error but prevents violation of the prescribed clearance.

If the safety filter becomes inactive after some finite time $T_s$, then
\begin{equation}
    \Delta u(t)=0,
    \qquad
    t\geq T_s.
\end{equation}
The error dynamics subsequently reduce to the nominal exponentially stable
system. Hence,
\begin{equation}
    \lim_{t\rightarrow\infty}e_p(t)=0,
    \qquad
    \lim_{t\rightarrow\infty}e_v(t)=0.
\end{equation}
Therefore, temporary obstacle avoidance does not prevent eventual recovery of
the nominal reference trajectory once the safety intervention is no longer
required.

\subsection{Feasibility under Acceleration Constraints}

The safety guarantees of Theorem~\ref{thm:forward_invariance} require
feasibility of the quadratic program. Define the set of safe and admissible
accelerations as
\begin{equation}
\begin{split}
    \mathcal{U}_{\mathrm{s}}(p,v)
    :=
    \bigl\{
        u\in\mathcal{U}:
        a_j(p)^\top u
        \geq b_j(p,v),\;
        j\in\mathcal{J}_a
    \bigr\}.
\end{split}
\label{eq:safe_admissible_input_set}
\end{equation}
The quadratic program is feasible if and only if
\begin{equation}
    \mathcal{U}_{\mathrm{s}}(p,v)\neq\varnothing.
    \label{eq:qp_feasibility_condition}
\end{equation}

For a single active obstacle, feasibility can be checked explicitly. Let
\begin{equation}
    a_j
    =
    \begin{bmatrix}
        a_{j,1}&a_{j,2}
    \end{bmatrix}^{\top}.
\end{equation}
The maximum value of $a_j^\top u$ over the box-constrained input set is
\begin{equation}
    \Gamma_j(p)
    :=
    \sum_{\ell=1}^{2}
    \begin{cases}
        a_{j,\ell}u_{\max,\ell},
        &a_{j,\ell}\geq0,\\
        a_{j,\ell}u_{\min,\ell},
        &a_{j,\ell}<0.
    \end{cases}
    \label{eq:maximum_barrier_authority}
\end{equation}
Therefore, the single-obstacle constraint is feasible if and only if
\begin{equation}
    \Gamma_j(p)\geq b_j(p,v).
    \label{eq:single_obstacle_feasibility}
\end{equation}

Condition \eqref{eq:single_obstacle_feasibility} has a direct interpretation:
the maximum available acceleration away from the obstacle must be sufficiently
large to satisfy the barrier requirement. A large inward velocity, small
acceleration bound, or late activation of the obstacle constraint can violate
this condition.

For multiple active obstacles, individual feasibility of every constraint is
necessary but may not be sufficient because the required acceleration
directions can conflict. In this case, feasibility is determined by the
intersection in \eqref{eq:safe_admissible_input_set}, which can be checked
using a small linear program before solving the quadratic program.
The present analysis therefore adopts the following assumption.

\begin{assumption}
\label{ass:qp_feasibility}
Along the closed-loop trajectory, the active obstacle configuration and
acceleration limits satisfy
\begin{equation}
    \mathcal{U}_{\mathrm{s}}(p(t),v(t))
    \neq
    \varnothing
\end{equation}
for all $t\geq0$.
\end{assumption}

The safety guarantee is conditional on compatible initial
conditions, timely inclusion of the relevant obstacle constraints,
and continued QP feasibility. Recursive feasibility is not
established. The fallback used in the simulations defines the
applied input when feasibility fails, but does not extend the
safety guarantee to those cases.

This assumption is explicitly examined in the numerical study by varying the
acceleration limits and reference speed. When the feasible set becomes empty,
no controller respecting the prescribed acceleration bounds can satisfy all
barrier inequalities simultaneously. In practice, feasibility can be
improved by activating obstacle constraints earlier, reducing the reference
speed, or increasing the available acceleration.

\begin{figure*}[!t]
    \centering
    \subfloat[Single-obstacle environment.]
    {
        \includegraphics[width=0.35\textwidth]
        {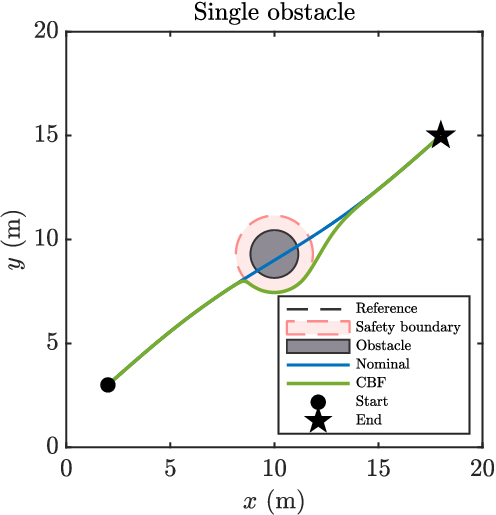}
        \label{fig:single_path}
    }
    \qquad
    \subfloat[Multiple-obstacle environment.]
    {
        \includegraphics[width=0.35\textwidth]
        {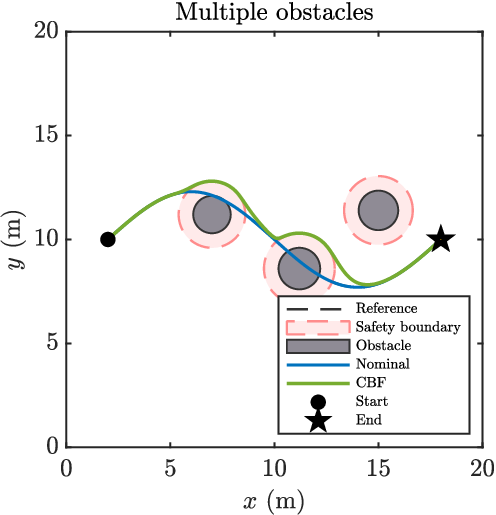}
        \label{fig:multiple_path}
    }
    \caption{Reference, nominal, and safety-filtered trajectories in the
    representative obstacle environments.}
    \label{fig:path_comparison}
\end{figure*}

\section{Numerical Experiments and Discussion}
\label{sec:numerical}

\subsection{Simulation Setup}

The proposed safety filter was evaluated in two planar environments containing
one and three circular obstacles, respectively. The controller gains and
barrier parameters were selected as
\begin{equation}
    K_p=3.2I_2,
    \qquad
    K_v=3.6I_2,
    \qquad
    \alpha_1=\alpha_2=2.
\end{equation}
The nominal gains yield the characteristic polynomial
$s^2+3.6s+3.2=(s+1.6)(s+2)$ in each coordinate, giving an
overdamped nominal response. The barrier gains yield
$\ddot h_j+4\dot h_j+4h_j\geq0$, whose associated equality
has a repeated pole at $-2$. These gains were fixed across
the reported tests and are not claimed to be optimal or
to guarantee continued QP feasibility.

The quadrotor radius and additional safety margin were
$r_{\mathrm{UAV}}=0.25$~m and $d_{\mathrm{safe}}=0.45$~m. Obstacles within
$3.5$~m of their effective boundary were included in the quadratic program.
To check this choice, let $\rho_j=\|p-p_{o,j}\|$, let $R_j$
denote the effective obstacle radius, and define the inward radial
velocity $v_{\mathrm{in},j}=-(p-p_{o,j})^\top v/\rho_j$.
Then
\[
    \psi_{j,1}\geq0
    \quad\Longleftrightarrow\quad
    v_{\mathrm{in},j}\leq
    \frac{\alpha_1(\rho_j^2-R_j^2)}{2\rho_j}.
\]
At $\rho_j=R_j+3.5$~m, $\alpha_1=2$ and
$R_j\in[1.60,1.85]$~m give admissible inward velocities
of approximately $4.60$--$4.71$~m/s.
Across the reported CBF simulations, the minimum $\psi_{j,1}$
at recorded points outside the activation region and upon
each entry into it was $28.7767>0$. This supports the activation
choice for the tested cases; continued QP feasibility remains
a separate assumption.

Unless otherwise stated, the acceleration limit was
$u_{\max}=2.5$~m/s$^2$.
The two-dimensional QP was solved by enumerating the nominal
input, its projections onto constraint boundaries, and pairwise
boundary intersections, then selecting the feasible candidate
closest to the nominal input. With $m$ active obstacles, there
are $M=m+4$ inequalities and at most $1+M+\binom{M}{2}$
candidates. Testing each candidate against all inequalities
gives $O(M^3)$ computational complexity, with at most 29
candidates in the three-obstacle case. Real-time execution
on onboard hardware was not assessed.

When the QP solver finds no feasible candidate, the simulation
applies the componentwise clipped nominal acceleration,
\[
    u_i=\min\{\max\{u_{\mathrm{nom},i},-u_{\max}\},u_{\max}\},
    \qquad i\in\{1,2\}.
\]
This fallback respects the acceleration bounds but does not
guarantee the barrier inequalities or obstacle avoidance.
The reported infeasible-step count records such events at
the stored simulation times, excluding internal Runge--Kutta
stage evaluations.

The closed-loop dynamics were integrated using a fourth-order
Runge--Kutta method with an integration step of $0.01$~s,
with the feedback recomputed at each Runge--Kutta stage.

For the single-obstacle environment, the reference trajectory was
\begin{equation}
    p_d(t)
    =
    \begin{bmatrix}
        2+16\xi\\
        3+12\xi+0.35\sin(2\pi\xi)
    \end{bmatrix},
    \qquad
    \xi=\frac{t}{20},
\end{equation}
and the obstacle was centered at $(10,9.3)$~m with radius $1.15$~m. For the
multiple-obstacle environment, the reference was
\begin{equation}
    p_d(t)
    =
    \begin{bmatrix}
        2+16\xi\\
        10+2.3\sin(2\pi\xi)
    \end{bmatrix},
    \qquad
    \xi=\frac{t}{22}.
\end{equation}
The three obstacle centers and radii were $(7,11.2,0.9)$,
$(11.2,8.6,1.0)$, and $(15,11.4,0.95)$, where the first two entries denote
the center coordinates in meters and the last entry denotes the physical
radius. In both cases, the quadrotor began at the initial reference position
with zero velocity.

The performance measures were the root-mean-square tracking error, maximum
tracking error, minimum safety clearance, control effort, command-modification
percentage, and number of infeasible quadratic programs. They were computed as
\begin{equation}
    e_{\mathrm{rms}}
    =
    \sqrt{
    \frac{1}{N}
    \sum_{k=1}^{N}
    \|p(t_k)-p_d(t_k)\|^2},
\end{equation}
\begin{equation}
    d_{\min}
    =
    \min_{j,k}
    \left(
    \|p(t_k)-p_{o,j}\|-d_j
    \right),
    \qquad
    J_u
    =
    \int_0^T\|u(t)\|^2\,\mbox{d}t.
\end{equation}
Thus, $d_{\min}\geq0$ indicates satisfaction of the prescribed safety
requirement. The modification percentage measures the fraction of samples at
which the applied acceleration differs from the nominal command due to the
safety filter or acceleration saturation.

\begin{figure*}[!t]
    \centering
    \subfloat[]{
        \includegraphics[width=0.24\textwidth]
        {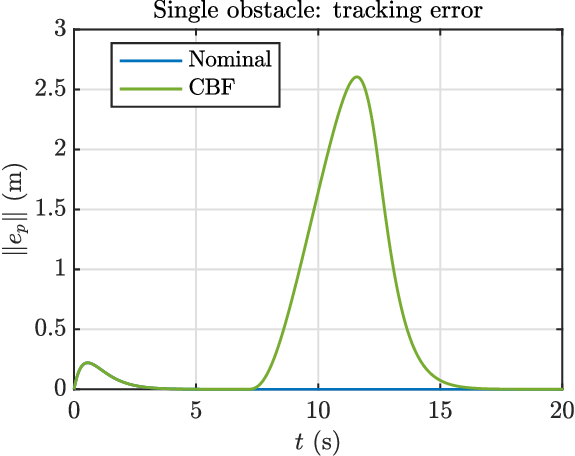}
        \label{fig:single_error}
    }
    \subfloat[]{
        \includegraphics[width=0.24\textwidth]
        {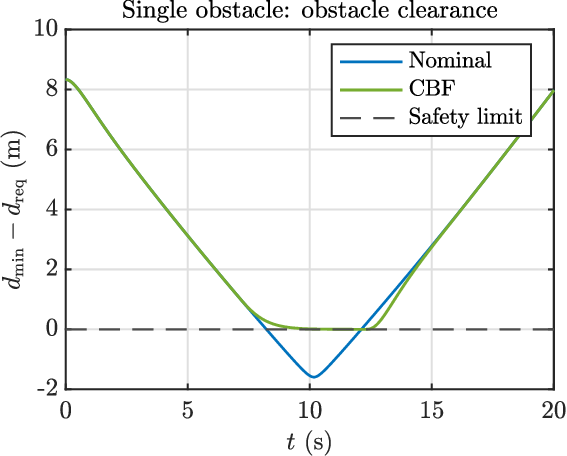}
        \label{fig:single_clearance}
    }
    \subfloat[]{
        \includegraphics[width=0.24\textwidth]
        {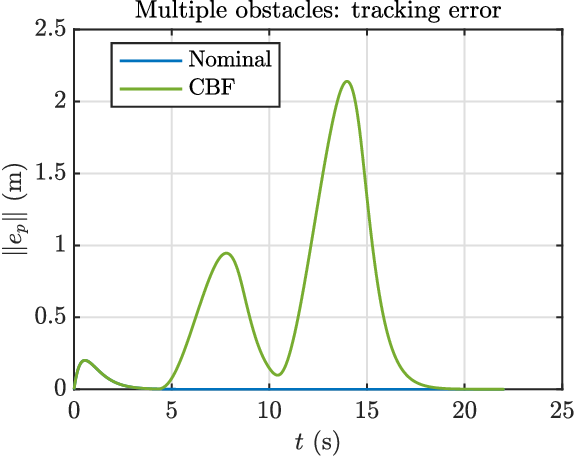}
        \label{fig:multiple_error}
    }
    \subfloat[]{
        \includegraphics[width=0.24\textwidth]
        {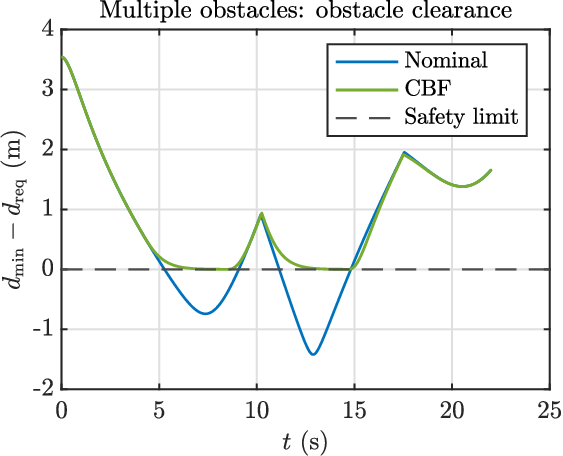}
        \label{fig:multiple_clearance}
    }
    \caption{Tracking errors and minimum obstacle clearances in the
    representative environments.
    (a) Tracking error with a single obstacle.
    (b) Clearance with a single obstacle.
    (c) Tracking error with multiple obstacles.
    (d) Clearance with multiple obstacles.}
    \label{fig:tracking_clearance}
\end{figure*}

\begin{table*}[!t]
    \centering
    \caption{Performance in the representative environments}
    \label{tab:fixed_scenarios}
    \resizebox{.975\textwidth}{!}{
    \begin{tabular}{llcccccc}
        \hline
        Scenario & Method
        & RMSE (m)
        & Maximum error (m)
        & $d_{\min}$ (m)
        & $J_u$
        & Modified (\%)
        & Infeasible steps\\
        \hline
        Single obstacle
        & Nominal & 0.050 & 0.221 & -1.594 & 2.559 & 0.300 & 0\\
        Single obstacle
        & CBF     & 0.908 & 2.605 &  0.000 & 10.893 & 26.237 & 0\\
        Multiple obstacles
        & Nominal & 0.044 & 0.201 & -1.421 & 2.539 & 0.091 & 0\\
        Multiple obstacles
        & CBF     & 0.747 & 2.140 &  0.001 & 9.759 & 40.800 & 0\\
        \hline
    \end{tabular}}
\end{table*}

\begin{figure*}[!t]
    \centering
    \subfloat[]{
        \includegraphics[width=0.24\textwidth]
        {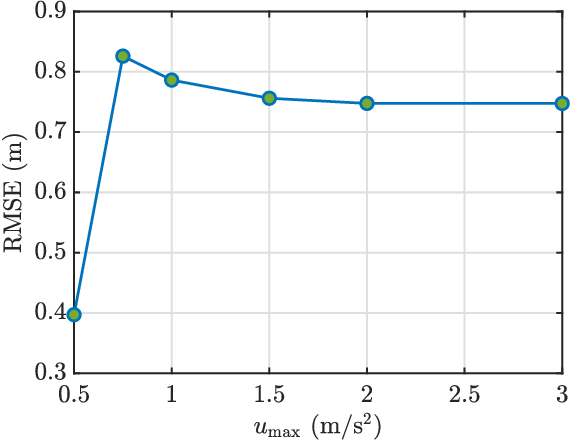}
        \label{fig:acceleration_rmse}
    }
    \subfloat[]{
        \includegraphics[width=0.24\textwidth]
        {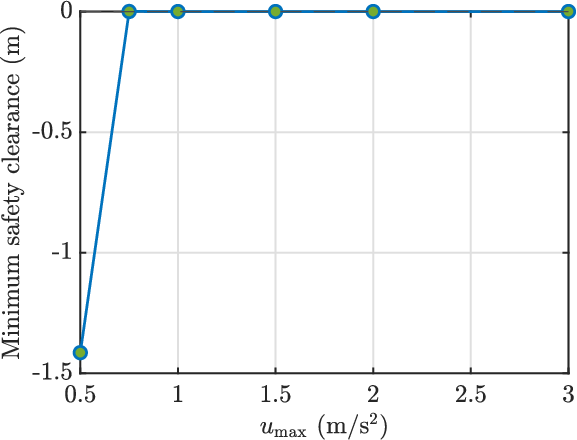}
        \label{fig:acceleration_clearance}
    }
    \subfloat[]{
        \includegraphics[width=0.24\textwidth]
        {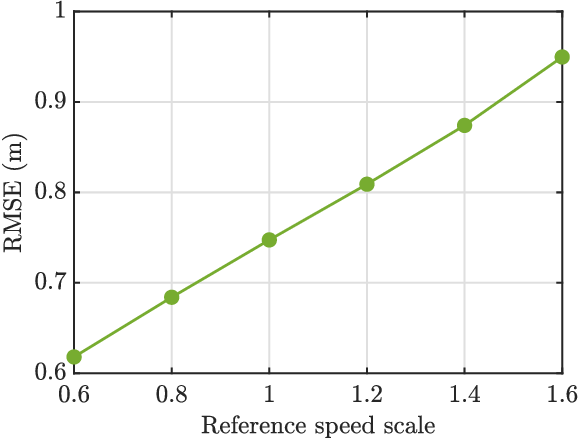}
        \label{fig:speed_rmse}
    }
    \subfloat[]{
        \includegraphics[width=0.24\textwidth]
        {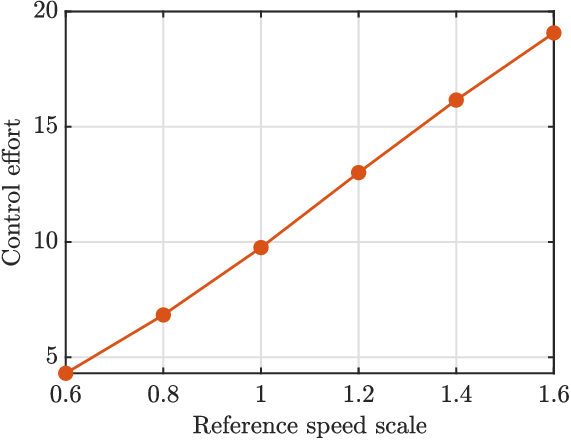}
        \label{fig:speed_effort}
    }
    \caption{Effects of acceleration authority and reference speed on the
    safety-filtered controller.
    (a) Acceleration limit and tracking RMSE.
    (b) Acceleration limit and minimum clearance.
    (c) Reference speed and tracking RMSE.
    (d) Reference speed and control effort.}
    \label{fig:acceleration_sensitivity}
\end{figure*}

\subsection{Single and Multiple Obstacle Environments}

Figure~\ref{fig:path_comparison} compares the nominal and safety-filtered
trajectories. In both environments, the nominal controller closely follows the
reference but enters the prescribed safety regions. The CBF controller departs
from the reference only near the obstacles, follows the corresponding safety
boundaries, and subsequently returns toward the desired trajectory.

The corresponding tracking errors and obstacle clearances are presented in
Figure~\ref{fig:tracking_clearance}. In the single-obstacle case, the tracking
error grows while the safety filter redirects the vehicle around the obstacle
and then converges toward zero after the intervention. A similar behavior is
observed in the multiple-obstacle case, where separate error peaks correspond
to successive avoidance maneuvers. In contrast, the nominal controller
maintains a smaller tracking error but produces negative clearance.

The quantitative results are summarized in
Table~\ref{tab:fixed_scenarios}. In the single-obstacle environment, the
nominal controller achieved an RMSE of $0.050$~m but violated the safety
requirement by $1.594$~m. The CBF controller maintained nonnegative clearance,
at the cost of increasing the RMSE to $0.908$~m and the control effort from
$2.559$ to $10.893$. Its command was modified during $26.237\%$ of the
simulation.

For the multiple-obstacle case, the nominal controller again produced a small
RMSE of $0.044$~m while violating the safety region by $1.421$~m. The proposed
controller maintained safety with an RMSE of $0.747$~m.
The applied command differed from the nominal command at $40.800\%$ of
recorded times due to barrier intervention or acceleration saturation.
No infeasible quadratic program occurred in either nominal test.
The clearances displayed as $0.000$ and $0.001$~m in the table are small
positive residuals rounded to three decimal places.

These results emphasize that accurate trajectory tracking does not imply safe
motion. The nominal controller reproduces the prescribed trajectory closely,
including the portions that intersect the obstacle regions. The safety filter
instead introduces only the deviation required to enforce the barrier
constraints.

\subsection{Effect of the Acceleration Limit}

The acceleration limit determines whether sufficient control authority is
available to satisfy the barrier constraints. The results of varying
$u_{\max}$ are shown in Figures~\ref{fig:acceleration_sensitivity} and
Table~\ref{tab:acceleration_sweep}.

\begin{table*}[!t]
    \centering
    \caption{Effect of the acceleration limit}
    \label{tab:acceleration_sweep}
    \resizebox{.75\textwidth}{!}{
    \begin{tabular}{ccccccc}
        \hline
        $u_{\max}$ (m/s$^2$)
        & RMSE (m)
        & $d_{\min}$ (m)
        & $J_u$
        & Modified (\%)
        & Infeasible steps
        & $\min\psi_1$\\
        \hline
        0.500 & 0.397 & -1.415 & 2.374 & 35.302 & 352 & -6.102\\
        0.750 & 0.826 &  0.001 & 6.125 & 57.156 &   0 &  0.000\\
        1.000 & 0.786 &  0.001 & 7.167 & 45.888 &   0 &  0.000\\
        1.500 & 0.756 &  0.001 & 8.723 & 42.572 &   0 &  0.000\\
        2.000 & 0.748 &  0.001 & 9.690 & 41.254 &   0 &  0.000\\
        3.000 & 0.747 &  0.001 & 9.762 & 40.709 &   0 &  0.000\\
        \hline
    \end{tabular}}
\end{table*}

\begin{table*}[!t]
    \centering
    \caption{Effect of the reference speed}
    \label{tab:speed_sweep}
    \resizebox{.75\textwidth}{!}{
    \begin{tabular}{ccccccc}
        \hline
        Speed scale
        & RMSE (m)
        & $d_{\min}$ (m)
        & $J_u$
        & Modified (\%)
        & Infeasible steps
        & $\min\psi_1$\\
        \hline
        0.600 & 0.618 & 0.000 &  4.306 & 35.915 & 0 & 0.000\\
        0.800 & 0.684 & 0.000 &  6.836 & 38.168 & 0 & 0.000\\
        1.000 & 0.747 & 0.001 &  9.759 & 40.800 & 0 & 0.000\\
        1.200 & 0.809 & 0.002 & 13.011 & 43.893 & 0 & 0.001\\
        1.400 & 0.874 & 0.004 & 16.159 & 47.837 & 0 & 0.003\\
        1.600 & 0.950 & 0.008 & 19.071 & 52.471 & 0 & 0.006\\
        \hline
    \end{tabular}}
\end{table*}

At $u_{\max}=0.500$~m/s$^2$, the quadratic program was infeasible at
$352$ sampled time instants. Consequently, the minimum clearance decreased to
$-1.415$~m and the minimum high-order barrier quantity became $-6.102$.
Although this case produced the smallest RMSE, the result is not desirable
because the vehicle failed to avoid the obstacles.

Increasing the limit to $0.750$~m/s$^2$ restored feasibility and produced
nonnegative clearance. The RMSE initially increased to $0.826$~m because the
controller now performed the required avoidance maneuvers. Further increases
in $u_{\max}$ reduced the RMSE and the command-modification percentage. Beyond
approximately $2.000$~m/s$^2$, the results changed only slightly, indicating
that the required barrier corrections no longer reached the imposed
acceleration bound.

This experiment confirms the feasibility qualification in the safety
analysis. The barrier condition cannot guarantee forward invariance when the
available acceleration is insufficient to satisfy the corresponding affine
constraints. It also demonstrates why tracking error alone is not an
appropriate performance measure for a safety-critical controller: the unsafe
case at $u_{\max}=0.500$~m/s$^2$ has a smaller RMSE because it does not execute
the necessary avoidance motion.

\subsection{Effect of the Reference Speed}

The reference trajectory was next traversed using speed scales from $0.600$ to
$1.600$, while the geometric path and controller parameters were unchanged.
The results are reported in Table~\ref{tab:speed_sweep} and
Figures~\ref{fig:speed_rmse}--\ref{fig:speed_effort}.

All tested reference speeds produced feasible quadratic programs and
nonnegative obstacle clearance. However, increasing the speed scale from
$0.600$ to $1.600$ increased the RMSE from $0.618$ to $0.950$~m and the
control effort from $4.306$ to $19.071$. The command-modification percentage
also increased from $35.915\%$ to $52.471\%$. Therefore,
a faster reference requires more frequent command modification and greater
control effort, leaving less
time for the vehicle to return toward the desired trajectory between
successive obstacles.

\begin{table*}[!t]
    \centering
    \caption{Gain sensitivity in the multiple-obstacle environment
    at $u_{\max}=2.5$~m/s$^2$ and speed scale $1$.}
    \label{tab:gain_sensitivity}
    \resizebox{.65\textwidth}{!}{
    \begin{tabular}{lccccc}
        \hline
        Setting & $K_p$ & $K_v$ & $\alpha_1=\alpha_2$
        & RMSE (m) & $d_{\min}$ (m)\\
        \hline
        Baseline & $3.2I_2$ & $3.6I_2$ & $2.0$ & 0.7474 & 0.000592\\
        Lower controller gains & $2.56I_2$ & $2.88I_2$ & $2.0$
            & 0.7674 & 0.000618\\
        Higher controller gains & $3.84I_2$ & $4.32I_2$ & $2.0$
            & 0.7345 & 0.000572\\
        Lower barrier gains & $3.2I_2$ & $3.6I_2$ & $1.6$
            & 0.7611 & 0.002330\\
        Higher barrier gains & $3.2I_2$ & $3.6I_2$ & $2.4$
            & 0.7412 & 0.000155\\
        \hline
    \end{tabular}}
\end{table*}

To assess gain sensitivity, we varied the controller gains
together and the barrier gains together by $\pm20\%$ in the
multiple-obstacle environment, keeping the other parameters fixed.
Table~\ref{tab:gain_sensitivity} shows that all five cases had zero
infeasible recorded steps and positive sampled minimum clearance.
Higher gains modestly reduced RMSE, although the clearance remained
small and decreased for the higher barrier gains. Because the
reference intersects the enlarged obstacle regions, the nominal
controller's low RMSE corresponds to unsafe motion; avoiding the
obstacles necessarily increases tracking error.

Overall, the experiments demonstrate that the proposed safety filter preserves
the nominal controller when its command is admissible and modifies it when
obstacle clearance is threatened. The results also illustrate the fundamental
tradeoff between tracking accuracy, safety, reference speed, and available
control authority. Most importantly, they confirm that the safety guarantee
depends on continued feasibility of the acceleration-constrained quadratic
program, as established in Section~\ref{sec:analysis}.

\section{Conclusion and Future Work}
\label{sec:conclusion}

This paper studied barrier-based safety filtering for fixed-altitude
quadrotor flight in cluttered environments under componentwise
acceleration limits. The filter uses a quadratic program to combine
high-order control barrier constraints with acceleration bounds. It
preserves the nominal acceleration command whenever that command satisfies
both sets of constraints.

For the planar double-integrator model, the analysis established exponential
tracking when the filter is inactive, bounded tracking error during safety
intervention, and forward invariance of the safe set from compatible initial
conditions while the quadratic program remains feasible. Simulations with
single and multiple stationary obstacles illustrated the effects of
acceleration limits and reference speed on clearance, tracking error, and
control effort. Tight acceleration limits can make the barrier constraints
infeasible, and the minimum clearances of $0.000$--$0.001$~m reported in
Table~\ref{tab:fixed_scenarios} leave little margin for sensing errors or
actuation delays.

Future work will address moving obstacles and delays \cite{fw-1,fw-2,fw-3},
model uncertainty and sampled-data implementation \cite{fw-4,fw-5},
learning-based control \cite{fw-6,fw-7,fw-8},
and recursive feasibility \cite{fw-9}.

\newpage

\bibliographystyle{ieeetr}  
\bibliography{arXiv}

\end{document}